\documentstyle [12pt]{article}
\newcommand{\be}{\begin{equation}}
\newcommand{\ee}{\end{equation}}
\newcommand{\ra}{\rightarrow}

\begin{document}

\begin{titlepage}
\begin{flushright}
 IFUP-TH 22/94\\
 April 1994\\
 hep-th/9404172
\end{flushright}

\vspace{7mm}

\begin{center}

{\Large\bf  Black holes as quantum membranes:\\

\vspace{2mm}

path integral approach}

\vspace{12mm}
{\large Michele Maggiore}

\vspace{3mm}

I.N.F.N. and Dipartimento di Fisica dell'Universit\`{a},\\
piazza Torricelli 2, I-56100 Pisa, Italy.\\
\end{center}

\vspace{4mm}

\begin{quote}
\hspace*{5mm} {\bf Abstract.} We describe the horizon
of a quantum black hole in terms of a
dynamical surface which defines the boundary of space-time as seen by
external static observers, and we define a path integral in the
presence of this dynamical boundary.
Using renormalization group
arguments, we find that
the dynamics of the horizon is governed by the action of the
relativistic bosonic membrane.
{}From the thermodynamical properties of
this bosonic membrane we derive the entropy and
the temperature of black
holes, and we find agreement with the standard results. With this
formalism we can also discuss the corrections to the Hawking
temperature when the mass $M$  of the black hole approaches
the Planck mass $M_{\rm Pl}$.
When $M$ becomes as low as $(10-100) M_{\rm Pl}$ a phase transition
takes place and the specific heat of the black hole becomes positive.

\end{quote}
\end{titlepage}
\clearpage

In a recent paper~\cite{MM} we have put forward a proposal for a
quantum description of black holes. Following the work of
't~Hooft~\cite{tH} and of Susskind and coworkers~\cite{Sus},
we have considered the black hole horizon, as seen by an external
static observer, as a dynamical surface endowed with physical degrees
of freedom, and we have suggested that the dynamics of this surface is
governed by an action principle. The simplest action
describing  a relativistic bosonic membrane in 3+1 dimensions
is~\cite{memb}
\be
I_{\rm memb}=-{\cal T}\int d^3\xi\, \sqrt{-h}\, ,
\ee
where ${\cal T}$ is the membrane tension, and $\xi^i=(\tau,
\sigma_1,\sigma_2 )$ parametrizes the membrane world-volume;
$h$ is the determinant of the induced metric
$h_{ij}=g_{\mu\nu}\partial_i\zeta^{\mu}\partial_j\zeta^{\nu}$
and $x^{\mu}=\zeta^{\mu}(\xi )$ gives the
embedding of the membrane in 3+1 dimensional space-time.
The target space metric $g_{\mu\nu}$ is taken to be the
Schwarzschild, Reissner-Nordstrom, or Rindler metric. We limit our
considerations to non-rotating black holes.

Classically, the action (1) describes a membrane which moves in the
black hole background, approaching the horizon asymptotically. For
instance, in the case of the Rindler metric
$ds^2=-g^2z^2dt^2+dx^2+dy^2+dz^2$
we have found a classical
solution of the form
\be\label{2}
z(\tau )=\frac{z_0}{\cosh g\tau}\, ,
\ee
which approaches asymptotically the
classical horizon, $z=0$. We have discussed
the quantization of the action~(1) in a minisuperspace approximation,
and we have found
that,  for a given value of  mass and charge, the quantum state of a
black hole is not uniquely determined, but rather there exists a
quasi-continuum of levels corresponding to excitations of the
membrane. In the classical description the state of
a (non-rotating) black
hole is characterized only by the ``macroscopic" parameters $M,Q$,
and therefore a
 coarse graining over the  membrane levels  is implicit.
The membrane approach therefore
 provides a microscopic explanation
of the black hole  entropy.
This description also suggests a quantization
 of the area of the horizon,
in agreement with various heuristic arguments
existing in the literature~\cite{quant}.

In this Letter we examine the membrane description from the path
integral point of view. We discuss how the action~(1) emerges from a
renormalization group analysis, and we show how
 the thermodynamical properties of black
holes can be computed in our approach. We will find that the black
hole radiation can be understood in terms of fluctuations of the
horizon -- a point of view stressed in particular by York~\cite{Yo83}.

Let us consider the definition of the path integral in the presence of
black holes. From the point
of view of a fiducial observer\footnote{For a non-rotating black hole
a fiducial observer is defined as a static observer outside the
horizon. The precise definition of fiducial observers for rotating
black holes, together with a detailed description of the membrane
paradigm for classical black holes, can be found in ref.~\cite{TPM}.}
only the degrees of freedom outside the horizon can influence the
dynamics. So, in a path integral approach, we
would like to perform the integration only
over the field variables $g_{\mu\nu}(x)$ with $x$
outside the (apparent) horizon.
However, a complication arises immediately,
since the position of the horizon
is determined by the metric itself: when the metric fluctuates,
the horizon fluctuates (see
e.g.~\cite{Yo83,Com,MM4}), and as a consequence the number
of variables $g_{\mu\nu}(x)$ which we would like to include
as integration variables
in the path integral changes, and we are faced with the problem of
defining a path integral in which the number of integration
variables fluctuates. A possibility would be to proceed to a gran
canonical ensemble.\footnote{I thank Enore Guadagnini for this
remark.} However, we rather
proceed as follows.
First, we introduce a fixed spherical surface located at $r=r_+
+\epsilon$, where $r_+$ is the radius of the classical horizon and
$\epsilon$ is larger than the
typical quantum fluctuations of the horizon,
which are on the order of a few Planck lengths. If $r>r_++\epsilon$ the
corresponding variables $g_{\mu\nu}$ are
inserted as integration variables
in the path integral. The question is what to do in the shell
$r_+<r<r_++\epsilon$.

In general, when evaluating a path integral in field
theory, a crucial point is the identification of the physically relevant
variables, as opposed to fast varying variables which can be integrated
out, in the spirit of Wilson's renormalization group~\cite{Wil}.
In a non-perturbative regime, the physically relevant variables are very
different from the fields which appear in the Lagrangian. An obvious
example is provided by {\em QCD}
at length scales on the order of one fm,
where the  physical degrees of freedom  are
hadrons rather than free quarks and gluons, and the gluon field
between a quark-antiquark pair is squeezed
in a flux tube whose dynamics is
governed, at the effective level,
by the action of a non-critical relativistic
string. In our case, a similar situation arises because,
from the point of view of a fiducial observer, the region within
a few Planck lengths from the classical horizon is a region of
super-Planckian temperatures and very large fields.

It is natural to identify the ``collective variables"
which play the main role with the coordinates $\zeta^{\mu}(\theta ,
\phi ,t)$ which define the position of the horizon,
$x^{\mu}=\zeta^{\mu}$.
Of course, the variables $\zeta^{\mu}$ are uniquely fixed by the metric
$g_{\mu\nu}$.
So, the integration variables $g_{\mu\nu}(x)$, restricted by
the condition that $x$ lies
between the horizon, as determined by
$g_{\mu\nu}$ itself, and the surface $r=r_++\epsilon$,
 are decomposed into the physically relevant variables
$\zeta^{\mu}$ plus ``fast variables". The effective action for the
$\zeta^{\mu}$ is obtained integrating over the fast variables. While
such integration would be very hard to perform explicitly, the general form
of the effective action is fixed by invariance principles. The simplest
term which can arise is just the membrane action, eq.~(1), with a membrane
tension ${\cal T}$ which is in principle derivable from the underlying
theory (even if its actual calculation is a difficult non-perturbative
problem, analogous to the computation of the string tension in {\em QCD}).

Thus, we can write the partition function as
\be\label{path1}
Z=\int{\cal D}\zeta^{\mu}\int_{\Omega}{\cal D}g_{\mu\nu}\, e^{iI}\, ,
\ee
where  $\Omega$ is the
region of space-time outside the
sphere $r=r_++\epsilon$. The action $I$ is given by
\be
I=I_{\rm memb} + I_{\rm grav}\,,
\ee
where $I_{\rm memb}$ is the membrane action, eq.~(1) (plus
higher-order terms which will be in general generated by the
renormalization group procedure), and
$I_{\rm grav}$ is the gravitational action of the variables
$g_{\mu\nu}(x)$ with $x\in\Omega$. A crucial point is that,
since the region $\Omega$ has a boundary
$\partial\Omega$, we must consider
the gravitational action supplemented with the boundary
term~\cite{Yo72,GH},
\be\label{5}
I_{\rm grav}=-\frac{1}{16\pi}\int_{\Omega} d^4x\sqrt{-g}\, R
-\frac{1}{8\pi}\int_{\partial\Omega} d^3x\sqrt{-\gamma}\, K\, ,
\ee
where $K$ is the trace of the extrinsic curvature of the boundary
$\partial\Omega$, i.e. of the surface $r=r_++\epsilon$,
and $\gamma$ is the determinant of the induced metric
on this surface (not to be confused
with $h$ which denotes the determinant
of the induced metric on the dynamical surface
$x^{\mu}=\zeta^{\mu}$; to avoid confusions let us
also stress again that
$\partial\Omega$ is a fixed {\em mathematical} surface while
$x^{\mu}=\zeta^{\mu}(\xi )$ gives the position of a {\em physical},
fluctuating, surface).

As it stands, eq.~(\ref{path1})
is only formal, since we have not specified what we mean by
integration over $g_{\mu\nu}$. The problem of the definition of the
path integral over $g_{\mu\nu}$ has been first discussed in the
well-known papers by Gibbons and Hawking~\cite{GH} and
Hawking~\cite{Haw2}. However, the important quantum fluctuations
are  that in the region $r_+<r<r_++\epsilon$, which have
already been taken into account by the renormalization group
procedure. Thus, the difficult problems
connected with a proper definition of the path integral over the
metrics can be avoided:  to lowest order in the semiclassical
expansion, the variables
$g_{\mu\nu}$ at $r>r_++\epsilon$ can be
replaced by the classical metric $g_{\mu\nu}^{\rm cl}$.

The definition of the path integral is completed
performing a Wick rotation on the
membrane world-sheet, $\tau\ra i\tau$. (Note that
$\tau$ appears both in $d^4x$ and in $d^3x$ in eq.~(\ref{5}) so that
both terms pick a factor $i$, and $\gamma$ becomes the Euclidean
induced metric.) Performing the Wick rotation
only at this stage, we avoid the problems connected with the
definition of Euclidean quantum gravity. Thus,
at this order, we  write
\be\label{path2}
Z=\left( e^{- \mbox{\normalsize {\em I}}_{\rm grav} }
\int{\cal D}\zeta^{\mu}\,
e^{- \mbox{\normalsize {\em I}}_{\rm memb} }
\right) |_{g_{\mu\nu}=g_{\mu\nu}^{\rm cl}}\, .
\ee
The path integral in the above equation is
the partition function of  an Euclidean membrane theory.
Note that
in the classical background metric $g_{\mu\nu}^{\rm cl}$
 the curvature
outside the horizon is zero and only the extrinsic curvature term
contributes to $I_{\rm grav}$.

In order to compute the thermodynamic properties of black holes
in the membrane formalism we proceed as follows. First,
it is convenient to choose the
gauge $\zeta^0=\tau$ in
the action~(1); thus, for a black hole metric
 $\tau$ coincides with the (Euclidean) time of
asymptotic observers. The solution given in  eq.~(\ref{2}) refers to
this choice of gauge. Then
we consider the partition function over
fields periodic in $\tau$ with period $\beta_{\infty}$ and we
evaluate the partition function in saddle point.
 In the case of the Rindler
metric a solution of the classical equations of motion
is given by eq.~(\ref{2}), and in Euclidean space it becomes an
instanton of the form
\be\label{inst}
z(\tau )=\frac{z_0}{\cos g\tau}\, .
\ee
We see that this solution is periodic in $\tau$ with period
$\beta_{\infty}=2\pi /g$. So it contributes to the path integral
for the partition function of a canonical ensemble with temperature
at infinity
\be
T=\frac{g}{2\pi}\, ,
\ee
which is the well known result for the temperature
measured in Rindler space~\cite{Unr}.

Let us now discuss the  Schwarzschild black hole.
 The equation of motion of the membrane
before the rotation $\tau\ra i\tau$
is~\cite{MM}
\be\label{eqm}
r\ddot{r}+2(\alpha^2-\dot{r}^2)+\frac{r\alpha '}{2\alpha}
(\alpha^2-3\dot{r}^2)=0\, ,
\ee
where $\alpha =1-2M/r$ and $\alpha '=d\alpha /dr$.
In more geometrical terms the above equation can be written
$K\left[\zeta\right] =0$, where
$K\left[\zeta\right]$ is the trace of the extrinsic
curvature of the surface $x^{\mu}=\zeta^{\mu}(\xi )$.
We define
\be
z=4M(1-\frac{2M}{r})^{1/2}\, .
\ee
This definition is motivated by the fact that,
with this change of
variables, together with $x=2M(\theta -(\pi /2))$ and
$y =2M\phi$, the Schwarzschild metric for a black hole with large
mass  goes into the Rindler metric,
plus corrections of order $O(z^2/M^2), O(x^2/M^2)$,
with the constant $g$ which appears
in the Rindler metric identified with the surface gravity
$\kappa =1/(4M)$. For the sake of comparison with the Rindler limit,
in the following we
will denote   the surface gravity by $g$. Note that
$0< z<1/g$ as $2M<r<\infty$. In terms of $z$
the equation of motion~(\ref{eqm}) reads
\be\label{11}
z\ddot{z}-2\dot{z}^2+g^2z^2(1+3g^2z^2)(1-g^2z^2)^3=0\, .
\ee
If $g\ra 0$ this equation reduces to the equation of
motion of the membrane in Rindler space,
$z\ddot{z}-2\dot{z}^2+g^2z^2=0$, which has the solution
given in eq.~(\ref{2}). This means that, if the mass of the black hole
$M$ goes to infinity, then $g=1/(4M)\ra 0$ and  equation~(\ref{11})
 (after the Wick rotation  $\tau\ra i\tau$) admits  a periodic solution
with period $2\pi /g$. Thus, for black holes in the limit $M\ra\infty$,
we recover the Hawking temperature~\cite{Haw74},
\be
T=\frac{1}{8\pi M}\, .
\ee
Beside reproducing the Hawking temperature in the large mass limit,
eq.~(\ref{11}) allows us to study how the result is modified at finite
mass. The semiclassical
approach used in the standard computation of black hole radiation
is valid only  if $M\gg M_{\rm Pl}$, where $M_{\rm Pl}$ is the Planck
mass, set equal to one in our units. Thus, in general, finite mass
corrections should be expected. The situation, however, is more
complicated. Eq.~(\ref{11}) cannot be studied expanding perturbatively
$z(\tau )=z_0(\tau )+g^2z_2(\tau )+g^4z_4(\tau )+
\ldots$ because $z_0(\tau )=a/\cos (g\tau)$
diverges at $g\tau =\pi /2$ and therefore the non-linear terms in
eq.~(\ref{11}) are not small, even if $g\ll 1$ (unless $g$ is
strictly zero).

To study eq.~(\ref{11}) we
introduce the function $w(\tau )=gz(\tau )$ and the variable
$t=g\tau$. In Euclidean space, the equation for $w(t)$ is
\be\label{12}
w\frac{d^2w}{dt^2}-2(\frac{dw}{dt})^2-w^2(1+3w^2)(1-w^2)^3=0\, .
\ee
By definition, $0<w<1$
since $0<z<1/g$. Apparently,
all dependence on $g$ has disappeared. However, when solving
eq.~(\ref{12}) with the initial conditions $w(0)=a,w'(0)=0$, where $a$
is a constant, $0<a<1$, we will find periodic solutions, see below.
The period of these
solution in general
depends on the amplitude $a$, because of the non-linearity
of the equation. Thus, one should ask what fixes the typical scale of
$a$. In the path integral, $a$ appears as a collective coordinate of
the instanton solution. In principle, the distribution of
values of $a$ is fixed
by the full theory, similarly to what happens for the instanton radius
in gauge theories. Qualitatively,
it is clear that the typical fluctuations of the horizon are of
order one in Planck units, so the main contribution comes from
solutions with $z(0)\sim 1$ and therefore $a\sim g$.
Fluctuations of the horizon with larger values of $a$
should be suppressed, and anyhow cannot be  computed in
the membrane formalism because,
as we have discussed in ref.~\cite{MM}, fluctuations with
amplitude $a\gg 1$ correspond to highly excited state of
the membrane, for which our description is not  self-consistent since
we are neglecting the backreaction of the membrane on the metric or,
equivalently, we are neglecting higher order terms in the membrane
effective action.

Thus, the dependence on $a$
that we will find below can be translated, apart
from numerical constants, into a dependence on $g$
just identifying $a\sim g$. In this way it is
possible to discuss, at least qualitatively,
finite mass corrections to the Hawking
temperature.

We have  integrated numerically eq.~(\ref{12}) for different
values of $a$. Let us introduce
the function $u(t)=1/w(t)$.
The results are as follows. For very small $a$
($a<0.004$) we find a periodic solution $u(t)$ which is essentially
undistinguishable, within our numerical accuracy, from the Rindler
limit with the same initial conditions, {\em i.e.} from the function
$(1/a)\cos t$. At $a=0.004$ we can put an upper bound
on the difference between the period $P$ of this solution
and the  period $P_0=2\pi$ of the unperturbed solution,
 $\Delta P/P_0 < 10^{-3}$.

Identifying  the period of the solution with
the inverse temperature of a black hole with
mass $M$ such that $g=1/(4M)\sim a$, we see that,
even for black holes with mass as low as $1/(4M)\sim 0.004$,
or $M\sim (10-100) M_{\rm Pl}$, the corrections to the Hawking
temperature are numerically neglegible --~if they exist at all.

For larger values of $a$ a  new phenomenon takes place.
An instability occurs, and the solution, instead of a form
qualitatively similar to
$(1/a)\cos t$, takes a form qualitatively similar to $(1/a)|\cos t|$,
{\em i.e.},
the negative part of the solution is flipped to
positive values. For  $a$ close to this critical value
 $a_c\simeq 0.004$, the
numerical integration becomes unreliable, with some half periods
flipped and others not flipped, randomly. The reality of the
phenomenon is however clearly seen increasing $a$ further.
Then, in correspondence with the
 point $\bar{t}\simeq \pi /2$
at which the derivative changes sign, the value of
$u(\bar{t})$ moves away from zero and instead of a cusp
the function $u(t)$ has zero derivative in $t=\bar{t}$, and the
numerical integration becomes stable again. The solution has now the
form shown in fig.~1.
The important new aspect is that this solution
has  a period which is one half as before,
corresponding to a temperature which is twice as
large as the Hawking temperature.

This result shows the existence of a non-analyticity in $g$. A
conservative interpretation of this non-analyticity is that it fixes
the limits of validity of the membrane description, or more in general
of the semiclassical approach. Certainly this is not surprising since we
know that the semiclassical expansion must break down when $M$
is not large compared to one. We find that this happens when
$1/(4M)\sim 0.004$, or $M\sim 10-100$, a reasonable result.

Of course, it is very tempting to see what
our equation predicts  for even larger
values of $a$, corresponding to an even smaller
value of the mass --~with the
remark that it is by no means obvious
that our results can be estrapolated
beyond this point of non-analyticity. We see that the equation suggests
that, at the critical point, the black hole still behaves as an
approximately thermal object, with a temperature twice as large as the
Hawking temperature. Increasing $a$ further, the period increases
smoothly (see fig.~2) and goes to infinity as $a\ra 1$, corresponding
to a black hole temperature which, after jumping from $1/(8\pi M)$
to $1/(4\pi M)$, starts to decrease
smoothly, until it becomes zero when
the mass is on the order of the Planck mass. A numerical fit for the
period $P$ as a function of $a$ shows that, when $a\ra 1$,
the data are very well
reproduced by $P={\rm const}\times (1-a)^{-1}$; identifying again
$a\simeq c g$, with $c\sim 1$, we get an expression for the temperature
 when $M$ becomes on the order of $M_{\rm Pl}$,
\be
T=Ag(1- c g)\, ,
\ee
where numerically
we find $A\simeq 1.5$. The temperature formally vanishes for
$g=1/c\sim 1$.

Independently of the numerical details,
the important new qualitative feature is that, for masses below a
critical value, the specific heat becomes positive. At the critical
point, a second order phase transition has taken place.

Although quite intriguing, the scenario for $M$ below the critical
value at present is only
speculative and we do not discuss it further here. Let us come back to
the safer domain of large black hole masses and discuss the
black hole entropy.

Having fixed
$\beta_{\infty}$ at the standard value, we can now compute
the entropy of black
holes, evaluating the partition function $Z$, and therefore the free
energy.  The leading term is given by the extrinsic
curvature term evaluated at the boundary $\partial\Omega$.
At this point the computation is formally identical
to a computation already performed by York~\cite{Yo86}:
the trace of the extrinsic curvature of a static
spherical surface with a generic radius $r$
in the   Schwarzschild background is
\be
K=-\frac{2}{r}\alpha^{1/2}-\frac{M}{r^2}\alpha^{-1/2}\, ,
\ee
The square root of the determinant
of the induced metric is
$\sqrt{\gamma} =r^2\alpha^{1/2}\sin\theta$ so that
\be
I_{\rm grav}= 12 \pi M^2 -8\pi Mr\, .
\ee
The local inverse temperature $\beta$ is defined as
\be\label{beta}
\beta =\beta_{\infty}\alpha^{1/2}=8\pi M
\left( 1-\frac{2M}{r}\right)^{1/2}\, .
\ee
Considering $I_{\rm grav}$ as a function of $r$
and $\beta$, with $M=M(r,\beta )$
defined by eq.~(\ref{beta}), one finds~\cite{Yo86} the entropy
\be
S=\beta\left( \frac{\partial I_{\rm grav}}
{\partial\beta}\right)-
I_{\rm grav}
=4\pi M^2\, ,
\ee
which agrees with the  Bekenstein-Hawking value.
It should be emphasized that this result is independent of $r$,
the radius
of the surface on which the extrinsic curvature is computed. This
explains the agreement between the value of the entropy  in the
membrane approach and the result of the computation
in Euclidean quantum gravity~\cite{GH,Haw2,Yo86}. In
the latter case
the boundary term is computed on a surface at infinity, while in our
approach it is computed at $r=r_++\epsilon$.
However, since the result is independent of $r$,
the two methods give the
same answer. The fact that the result is independent of $r$
has also the important consequence that the dependence
on the exact position of the surface $\partial\Omega$ disappears.

It is interesting to compare this derivation of the black hole entropy
with a recent result of Frolov and Novikov~\cite{FN}. There is an
important conceptual difference, since in ref.~\cite{FN}
the entropy is
attributed to the modes of physical fields which are located
{\em inside} the
horizon and therefore invisible for a distant observer. Instead in
our derivation, in the spirit of the principle of black hole
complementarity~\cite{Sus},
we have avoided any reference to the ``inside'' of the
black hole. However, we think that some similarities  exist
between the approach
presented in this paper  and the approach of ref.~\cite{FN}; in
both cases the entropy is obtained tracing over some degrees of
freedom. In our case, the extrinsic curvature term which is responsible
for  the entropy is generated by the renormalization group procedure,
which is a way of tracing over short scale fluctuations in the
vicinity of the horizon, while  in
\cite{FN} a distinction between ``visible'' and invisible'' modes is
made, and the entropy is obtained performing the appropriate
 trace over invisible
modes. It turns out that  the main contribution to the entropy is
given by the ``invisible'' modes which propagate in the narrow shell
$r_+<r<r_++\epsilon$, with $\epsilon$ on the order of a few Planck
length. So, if we rather define the
``invisible'' modes as the modes which propagate in this shell,
without making any reference to the inside of the black hole,
the result $S\sim A$ found in~\cite{FN} still goes through, but the
derivation is now conceptually very similar to what we have presented.
The proportionality constant between the entropy $S$ and the area $A$
can only be estimated, but not exactly computed, in the approach of
ref.~\cite{FN}. In the membrane approach, instead,  it can be computed
exactly, as we have seen.
The difference is due to the fact that we perform the ``tracing''
 at the level of the action, via the renormalization group.
This produces the extrinsic curvature term on the surface
$r=r_++\epsilon$, with a coefficient which
is uniquely fixed by the fact that
it must cancel the term with second derivatives in the
Einstein-Hilbert action.

We can also try to go further an compute
the corrections to the leading term in the entropy,
inserting the instanton solution into the membrane action
(and integrating
over the collective coordinates of the instanton).
Further corrections are obtained computing the fluctuations around the
instanton. However, in general this produces divergent terms.
For instance in
Rindler space, where the instanton has the form~(\ref{inst}),
the action of the solution is divergent,
\be
S_{\rm inst}={\cal T'}z_0\int_0^{2\pi}dt\,
\frac{1}{\cos^2t}\, ,
\ee
where ${\cal T'}={\cal T} \int dxdy$ is proportional to the area of
the membrane and the proportionality constant, given by the integral
in $dt$, diverges.
Thus, in general we have (writing explicitly also the Newton constant
$G$)
\be
S=\frac{1}{4G}(1+C)A
\ee
where $A$ is the area and $C$ a formally divergent constant.
This divergence does not come out as a surprise.
A similar  divergence has been found by 't~Hooft~\cite{tH}
computing  the number of states of a scalar field in the
vicinity of the horizon.  Its relation with the
ultraviolet problem in quantum gravity has been discussed
in~\cite{SU} where it has been suggested that this divergence is
reabsorbed by the renormalization of Newton's constant.
In the membrane approach, this divergence can be
traced to the fact that, in the semiclassical approximation,
the quantum membrane dynamics is governed by a potential
which admits a continuum spectrum~\cite{MM}.
However, this only happens if we extrapolate the membrane effective
potential to the region very close to the horizon, where
from the point of view of the fiducial observer the temperature
becomes super-Planckian and quantum gravity effects set in. Thus,
the divergence in the entropy is actually a short distance problem,
related to the fluctuations of the metric at the Planck scale
$L_{\rm Pl}$.  Instead, the
 membrane approach that we have presented, being derived from a
renormalization group procedure, can only give an effective
description valid at distances larger than $L_{\rm Pl}$.

\end{document}